\documentclass[aps,showpacs,preprintnumbers,amsmath,amssymb]{revtex4}
\usepackage{dcolumn}
\usepackage[dvips]{epsfig}

\usepackage{float}
\usepackage{graphics,epsfig}
\usepackage{graphicx}
\usepackage{dcolumn}
\usepackage{bm}
\usepackage{gensymb}
\usepackage{subfigure}

\begin{document}
\baselineskip=0.8 cm
\title{\bf The surface geometry and shadow of a Schwarzschild black hole with halo}

\author{Mingzhi Wang$^{1}$\footnote{wmz9085@126.com}, Guanghai Guo$^{1}$\footnote{Corresponding author: thphys\_qust@163.com}, Songbai Chen$^{2,3}$\footnote{csb3752@hunnu.edu.cn},  Jiliang Jing$^{2,3}$\footnote{jljing@hunnu.edu.cn}
}
\affiliation{$ ^1$School of Mathematics and Physics, Qingdao University of Science and Technology, Qingdao, Shandong 266061, People's Republic of
China \\ $ ^2$Institute of Physics and Department of Physics, Key Laboratory of Low Dimensional Quantum Structures
and Quantum Control of Ministry of Education, Synergetic Innovation Center for Quantum Effects and Applications,
Hunan Normal University, Changsha, Hunan 410081, People's Republic of China\\
$ ^3$Center for Gravitation and Cosmology, College of Physical Science and Technology,
Yangzhou University, Yangzhou 225009, China}

\begin{abstract}
\baselineskip=0.6 cm
\begin{center}
{\bf Abstract}
\end{center}
We have studied the surface geometry and shadows of Schwarzschild black hole with a halo containing the quadrupolar and octopolar terms. We found the quadrupole term makes the Schwarzschild black hole prolate for the quadrupole strength $\mathcal{Q}<0$ and oblate for $\mathcal{Q}>0$, and the octopole term makes shadow stretch upward for the octopolar strength $\mathcal{O}<0$ and stretch downward for $\mathcal{O}>0$. The shadow of Schwarzschild black hole with halo stretches and squeezes along the horizontal direction for $\mathcal{Q}<0$ and $\mathcal{Q}>0$ respectively. Meanwhile, black hole shadow shifts upward for $\mathcal{O}<0$ and shifts downward for $\mathcal{O}>0$. We exhibited the light rays that form the shadow boundary to explain the emergence of the extraordinary patterns of black hole shadow with the quadrupole and octopole terms. From the observable width $W$, height $H$, oblateness $K$ and distortion parameter $\delta_{c}$ of black hole shadow, one can determine the quadrupole and octopolar strengths of Schwarzschild black hole with halo. Our results show that the quadrupolar and octopolar terms yield a series of interesting patterns for the shadow of a Schwarzschild black hole with halo.

\end{abstract}

\pacs{ 04.70.Dy, 95.30.Sf, 97.60.Lf } \maketitle
\newpage
\section{Introduction}

Nowadays, Event Horizon Telescope (EHT) Collaboration et al have revealed the first images of the supermassive black hole at the centre of the giant elliptical galaxy M87\cite{eht,fbhs1,fbhs2,fbhs3,fbhs4,fbhs5,fbhs6} and the Milky Way Galaxy\cite{sga1,sga2,sga3,sga4,sga5,sga6}. It is of great help to the study of black hole physics and astrophysics, attracts more and more researchers to devote themselves to the research of black hole images. A very important element in black hole image is black hole shadow\cite{synge,sha2,lumi,sha3}, a dark silhouette. The dark shadow appears because the light rays close to the event horizon are captured by black hole, thereby leaving a black shadow in the observer's sky. Due to the fingerprints of the geometry around black hole could be reflected in the shape and size of shadow, the research of black hole shadows plays a vital role in the study of black holes (constraining black hole parameters)\cite{sha9,sha8,dressed,Intcur,bhparam,obsdep,constr}, probing some fundamental physics issues including dark matter\cite{polar7, drk, polar8,shadefl,shasgra} and verification of various gravity theories\cite{safeg,lf,sha10,fR, 2101, 2107, 2111}. It is a perfect black disk for Schwarzschild black hole shadow, and it gradually becomes a D-shaped silhouette with the increase of spin parameter for Kerr black hole shadow\cite{sha2}. In the space-time of a Kerr black hole with Proca hair and a Konoplya-Zhidenko rotating non-Kerr black hole, the cusp silhouette of black hole shadows emerge\cite{fpos2,sb10}. The self-similar fractal structures appear in the black hole shadow originating from the chaotic lensing\cite{sw,swo,astro,chaotic,my,sMN,zhengwen23,sha18,binary,bsk}. Many other black hole shadows with other parameters in various theories of gravity have been recently investigated in Refs. \cite{swo7,mbw,mgw,msr,mcs,schm,scc,sha4,sha5,sha6,sha7,sha11,sha111,sha12,sha13,sha14,who,whr,sha141,sha15,sha16,sb1,sha17,sha19,sha191,sha192,sha193,sha194,shan1,shan1add,shan3add,rr,pe,lf2,Zeng2020vsj,Zeng2020dco,lens,knn,bieu,ssp,nake,nakeop,BI}. It is hope that these information imprinted in black hole shadows can be captured in the future astronomical observations including the upgraded Event Horizon Telescope and BlackHoleCam\cite{bhc} to study black holes and verify various gravity theories.

It is widely believed that a massive halo, ring or other shell-like distributions of matter could be concentrated around black hole in the galactic center. The first image of a Schwarzschild black hole with
thin accretion disk was announced by Luminet\cite{lumi}, it shows the primary and secondary images of the thin accretion disk around shadow. The superposition of a black hole and exterior matter will reorganize the space-time structure, and bring a huge change to black hole shadow. We have researched the shadows of a Schwarzschild black hole surrounded by a Bach-Weyl ring\cite{mbw}, and found black hole shadow becomes a ``8" type shaped silhouette, and possesses self-similar fractal structures. P. V. P. Cunha et al \cite{lens} researched the shadows of a black hole surrounded by a heavy Lemos-Letelier accretion disk, and found black hole shadow becomes more prolate with the increase of the accretion disk mass. In this paper we consider a solution of the Einstein equations that represents the superposition of a Schwarzschild black hole with a halo containing the quadrupolar and octopolar contributions \cite{halo}. For this space-time, W. M. Vieira et al detected the timelike geodesic orbits of test particles, and found the octopolar term could bring about chaotic motions\cite{halo}. By further numerical search, they found the quadrupolar term also gives rise to a chaotic behavior in addendum\cite{Add}. F. L. Dubeibe further investigated the effects of the quadrupole and octupole moments on the motion of a test particle by using modern color-coded basin diagrams\cite{shp}, and found the final state of the test particle is highly affected. Therefore the nonlinear superposition of Schwarzschild black hole with halo also could influence the null geodesic motions of photons, and then affects the black hole shadow. In this paper, we will probe how the quadrupolar and octopolar terms affect Schwarzschild black hole shadow.

The paper is organized as follows. In section II, we briefly review the space-time of Schwarzschild black hole with halo, and reveal the influences of the quadrupolar and octopolar terms on the surface geometry of black hole. In Section III, we present numerically the shadows of Schwarzschild black hole with halo, and analyse the new features of black hole shadow arising from the quadrupolar and octopolar terms. Finally, we end the paper with a summary.

\section{The surface geometry of Schwarzschild black hole with halo}

The space-time of Schwarzschild black hole with halo is a vacuum static and axially symmetric space-time, so it can be described by the Weyl metric
\begin{eqnarray}
\label{zjzdg}
ds^{2}=-e^{2\nu}dt^{2}+e^{2\lambda-2\nu}
(d\rho^{2}+dz^{2})+\rho^{2}e^{-2\nu}d\phi^{2},
\end{eqnarray}
where $\nu$ and $\lambda$ only are the functions of $\rho$ and $z$. The Einstein equations reduce to
\begin{eqnarray}
\label{ee}
&&\nu_{,\rho\rho}+\frac{\nu_{,\rho}}{\rho}+\nu_{,zz}=0,\\
&&\lambda_{,\rho}=\rho(\nu_{,\rho})^{2}-\rho(\nu_{,z})^{2},\;\;\;\;\;\;\lambda_{,z}=2\rho\nu_{,\rho}\nu_{,z}.
\end{eqnarray}
The function $\nu(\rho, z)$ satisfies the Laplace equation and behaves like the gravitational potential in the Newtonian theory, thus it can be superposed linearly. However, the function $\lambda(\rho, z)$ does not own such a property of linear superposition. In the solution of Schwarzschild black hole with halo, the functions $\nu$ and $\lambda$ for the whole system can be written as $\nu=\nu_{Schw}+\nu_{halo}$ and $\lambda=\lambda_{Schw}+\lambda_{halo}+\lambda_{int}$, respectively\cite{ong}. The functions $\nu_{Schw}, \lambda_{Schw}$ are the solution of Schwarzschild space-time described as
\begin{eqnarray}
\label{schw}
&&\nu_{Schw}=\frac{1}{2}\ln\frac{d_{1}+d_{2}-2M}{d_{1}+d_{2}+2M}\\
&&\lambda_{Schw}=\frac{1}{2}\ln\frac{(d_{1}+d_{2})^{2}-4M^{2}}{4d_{1}d_{2}},
\end{eqnarray}
where $M$ is the mass of Schwarzschild black hole, and $d_{1,2}=\sqrt{\rho^{2}+(z\mp M)^{2}}$. The functions $\nu_{halo}, \lambda_{halo}$ are the solution of halo structure, and $\lambda_{int}$ represents the interaction between the black hole and halo.

In the Schwarzschild coordinates, the space-time of Schwarzschild black hole with halo can be described by the metric \cite{halo}
\begin{eqnarray}
\label{shdg}
ds^{2}=-(1-\frac{2M}{r})e^{(\nu_{Q}+\nu_{O})}dt^{2}+e^{(\lambda_{Q}+\lambda_{O}+\lambda_{QO}-\nu_{Q}-\nu_{O})}\bigg[(1-\frac{2M}{r})^{-1}dr^{2}+r^{2}d\theta^{2}\bigg]+e^{-(\nu_{Q}+\nu_{O})}r^{2}\sin^{2}\theta d\phi^{2},
\end{eqnarray}
where
\begin{eqnarray}
\label{halo}
\nu_{Q}&=&(\mathcal{Q}/3)(3u^{2}-1)(3v^{2}-1),\\ \nonumber
\nu_{O}&=&(\mathcal{O}/5)uv(5u^{2}-3)(5v^{2}-3),\\ \nonumber
\lambda_{Q}&=&-4\mathcal{Q}u(1-v^{2})+(\mathcal{Q}^{2}/2)[9u^{4}v^{4}-10u^{4}v^{2}-10u^{2}v^{4}+12u^{2}v^{2}+u^{4}+v^{4}-2u^{2}-2v^{2}+1],\\ \nonumber
\lambda_{O}&=&-2\mathcal{O}[3u^{2}v-3u^{2}v^{3}+v^{3}-\frac{9}{5}v+\frac{4}{5}]+2\mathcal{O}^{2}[\frac{75}{8}u^{6}v^{6}-\frac{117}{8}u^{6}v^{4}-\frac{117}{8}u^{4}v^{6}+\frac{45}{8}u^{6}v^{2}+\frac{45}{8}u^{2}v^{6}\\ \nonumber &+&\frac{189}{8}u^{4}v^{4}-\frac{387}{40}u^{4}v^{2}-\frac{387}{40}u^{2}v^{4}+\frac{891}{200}u^{2}v^{2}-\frac{3}{8}u^{6}-\frac{3}{8}v^{6}+\frac{27}{40}u^{4}+\frac{27}{40}v^{4}-\frac{81}{200}u^{2}-\frac{81}{200}v^{2}+\frac{21}{200}],\\ \nonumber
\lambda_{QO}&=&2\mathcal{Q}\mathcal{O}[9u^{5}v^{5}-12u^{5}v^{3}+3u^{5}v-12u^{3}v^{5}+\frac{84}{5}u^{3}v^{3}-\frac{24}{5}u^{3}v+3uv^{5}-\frac{24}{5}uv^{3}+\frac{9}{5}uv],
\end{eqnarray}
and
\begin{eqnarray}
\label{uv}
&&u=\frac{1}{2M}\bigg[\sqrt{\rho^{2}+(z+M)^{2}}+\sqrt{\rho^{2}+(z-M)^{2}}\bigg],\\ \nonumber
&&v=\frac{1}{2M}\bigg[\sqrt{\rho^{2}+(z+M)^{2}}-\sqrt{\rho^{2}+(z-M)^{2}}\bigg].
\end{eqnarray}
The transformation between the Schwarzschild coordinates ($r, \theta$) and the Weyl coordinates ($\rho, z$) is:
\begin{eqnarray}
\label{bh}
\rho=\sqrt{r(r-2M)}\sin\theta,\;\;\;\;\;\;\;\;z=(r-M)\cos\theta.
\end{eqnarray}
The exterior halo is a multipolar structure containing quadrupolar and octopolar terms, and $\mathcal{Q}$ and $\mathcal{O}$ are the quadrupole and octopole strengths respectively. The metric will reduce to Schwarzschild solution when $\mathcal{Q}=\mathcal{O}=0$.

\begin{figure}
\includegraphics[width=16.5cm ]{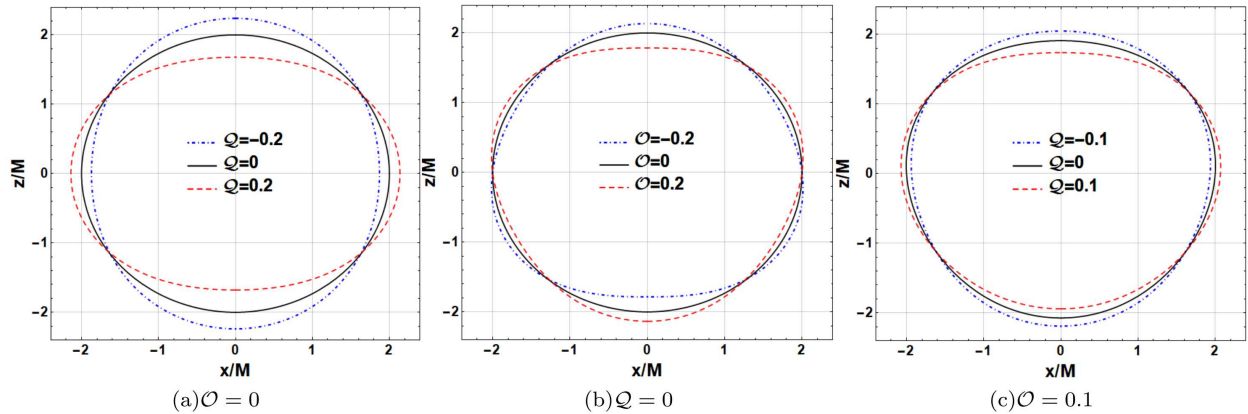}
\caption{The embedding diagrams for the surface geometry of Schwarzschild black hole with halo in Euclidean 3-space with the different quadrupole strength $\mathcal{Q}$ and octopole strength $\mathcal{O}$.}
\label{ehqo}
\end{figure}

The existence of halo does not change the radial coordinate of the event horizon, $r_{h}=2M$, but the surface geometry of Schwarzschild black hole with halo would be affected by the quadrupole strength $\mathcal{Q}$ and octopole strength $\mathcal{O}$. The event horizon of Schwarzschild black hole with halo can be described by the two-dimensional line element
\begin{eqnarray}
\label{ehdg}
ds_{h}^{2}=r_{h}^{2}e^{(-\nu_{Q}-\nu_{O})}|_{r=r_{h}}(d\theta^{2}+\sin^{2}\theta d\phi^{2}).
\end{eqnarray}
The embedding diagrams \cite{sgcr,sgmf} for the event horizon of Schwarzschild black hole with halo in Euclidean 3-space with the different $\mathcal{Q}$ and $\mathcal{O}$ are exhibited in Fig.\ref{ehqo}. Since the event horizon in plane ($x,y$) is centrosymmetric, we only show the event horizon in plane ($x,z$), where $z=0$ represents the equatorial plane. From Fig.\ref{ehqo}(a) with $\mathcal{O}=0$, one can find the event horizon is prolate for the quadrupole strength $\mathcal{Q}<0$, and it could become more prolate by stretching along the vertical direction and squeezing along the horizontal direction as $\mathcal{Q}$ decreases. But for $\mathcal{Q}>0$, the event horizon becomes more oblate in the opposite way as $\mathcal{Q}$ increases. From Fig.\ref{ehqo}(b) with $\mathcal{Q}=0$, one can find the octopolar term ($\mathcal{O}\neq0$) breaks the reflection symmetry of event horizon with respect equatorial plane, but the event horizon in the equatorial plane doesn't change with $\mathcal{O}$. The event horizon stretches upward for $\mathcal{O}<0$ and stretches downward for $\mathcal{O}>0$. The two event horizons with opposite $\mathcal{O}$ are symmetrical to each other about the equatorial plane. Fig.\ref{ehqo}(c) shows the joint efforts of the quadrupole strength $\mathcal{Q}$ and the octopole strength $\mathcal{O}$ on the event horizon for $\mathcal{O}=0.1$ and $\mathcal{Q}=-0.1,0,0.1$. That is the quadrupole term makes Schwarzschild black hole prolate for $\mathcal{Q}<0$ and oblate for $\mathcal{Q}>0$, and the octopole terms make shadow stretch upward for $\mathcal{O}<0$ and stretch downward for $\mathcal{O}>0$.

\begin{figure}
\includegraphics[width=15cm ]{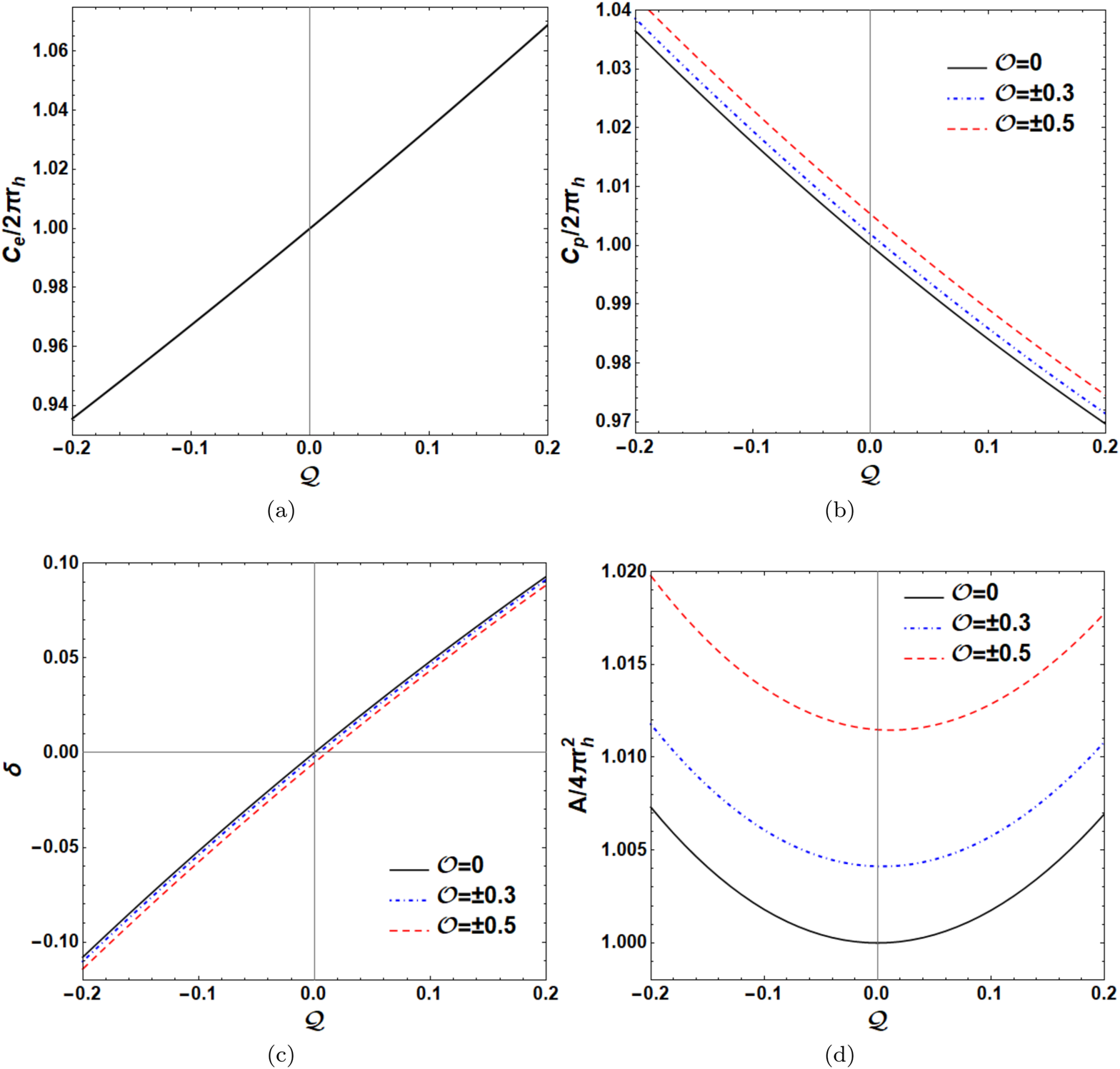}
\caption{The changes of the equatorial circumference $C_{e}$, the polar circumference $C_{p}$, the distortion parameter $\delta$ and the surface area $A$ of the event horizon with the quadrupole strength $\mathcal{Q}$ for different octopole strength $\mathcal{O}$. Here we set Schwarzschild black hole mass $M=1$.}
\label{ca}
\end{figure}

To illustrate the influences of the quadrupolar and octopolar terms on Schwarzschild black hole with halo quantitatively, we compute the equatorial circumference $C_{e}$ and the polar circumference $C_{p}$ of the event horizon, which are given by
\begin{eqnarray}
\label{cep}
C_{e}&=&\int^{2\pi}_{0}\sqrt{g_{\phi\phi}}d\phi\bigg|_{(r=r_{h},\theta=\frac{\pi}{2})}=\int^{2\pi}_{0}r_{h}e^{-\frac{1}{2}(\nu_{Q}+\nu_{O})}|_{(r=r_{h},\theta=\frac{\pi}{2})}d\phi,\\
C_{p}&=&2\int^{\pi}_{0}\sqrt{g_{\theta\theta}}d\theta\bigg|_{r=r_{h}}=2\int^{\pi}_{0}r_{h}e^{-\frac{1}{2}(\nu_{Q}+\nu_{O})}|_{r=r_{h}}d\theta.
\end{eqnarray}
Fig.\ref{ca}(a) and (b) show the plots of the equatorial circumference $C_{e}$ and the polar circumference $C_{p}$ of the event horizon as a function of $\mathcal{Q}$. Where $2\pi r_{h}$ is the circumference of Schwarzschild black hole event horizon ($\mathcal{Q}=\mathcal{O}=0$). With the increase of $\mathcal{Q}$, one can find $C_{e}$ increases and $C_{p}$ decreases. The equatorial circumference $C_{e}$ is independent of $\mathcal{O}$, which are consistent with what is shown in Fig.\ref{ehqo}(b). The two polar circumferences $C_{p}$ with opposite $\mathcal{O}$ share the same value, and increase as $|\mathcal{O}|$ increase for fixed $\mathcal{Q}$. In addition, one can find the change of $C_{p}$ with $\mathcal{O}$ is very small, it signifies the octopole strength $\mathcal{O}$ has little effect on the polar circumferences $C_{p}$. The distortion parameter $\delta$ defined as $(C_{e}-C_{p})/C_{e}$ can determine the departure of the event horizon from spherical symmetry. Fig.\ref{ca}(c) shows the behaviors of the distortion parameter $\delta$ with $\mathcal{Q}$ and $\mathcal{O}$. The distortion parameter $\delta$ almost depends only on $\mathcal{Q}$, and it nearly is less than $0$ for $\mathcal{Q}<0$, meaning the prolate event horizon, $\delta>0$ means the oblate event horizon. The surface area $A$ of the event horizon is given by
\begin{eqnarray}
\label{A}
A=\int^{\pi}_{0}\int^{2\pi}_{0}\sqrt{g_{\phi\phi}g_{\theta\theta}}d\theta d\phi\bigg|_{r=r_{h}}=\int^{\pi}_{0}\int^{2\pi}_{0}r_{h}^{2}\sin\theta e^{-(\nu_{Q}+\nu_{O})}|_{r=r_{h}}d\theta d\phi.
\end{eqnarray}
Fig.\ref{ca}(d) shows the change of the surface area $A$ of the event horizon with $\mathcal{Q}$ and $\mathcal{O}$, where $4\pi r_{h}^{2}$ is the surface area of Schwarzschild black hole. One can find the surface area $A$ first decreases and then increases with the increase of $\mathcal{Q}$, which means whether the prolate or oblate event horizon both have bigger surface area. The increase of surface area $A$ with $|\mathcal{O}|$ means the octopole term also can enlarge the surface area of schwarzschild black hole.

\begin{figure}
\includegraphics[width=8cm ]{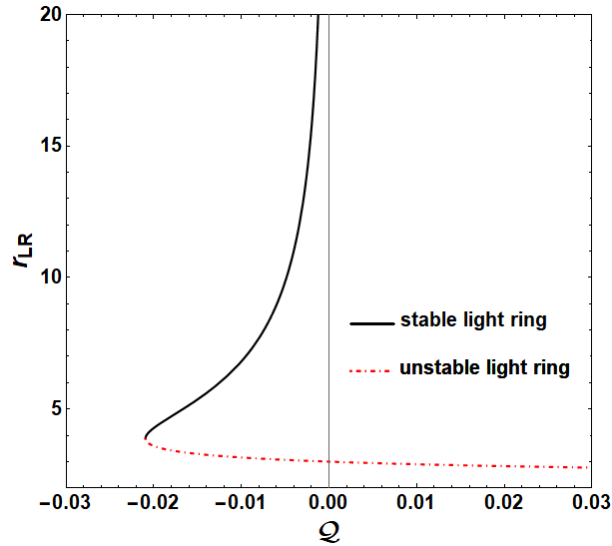}
\caption{The plots of the radiuses $r_{LR}$ of unstable (red dashed line) and stable (black line) light rings with the quadrupole strength $\mathcal{Q}$.}
\label{rlq}
\end{figure}

The Hamiltonian $\mathcal{H}$ of a photon propagation along null geodesics in the space-time of Schwarzschild black hole with halo (\ref{shdg}) can be described as:
\begin{eqnarray}
\label{hmd}
\mathcal{H}=g^{rr}p_{r}^{2}+g^{\theta\theta}p_{\theta}^{2}+V_{eff}=0,
\end{eqnarray}
where the effective potential $V_{eff}$ is defined as
\begin{eqnarray}
\label{veff}
V_{eff}=g^{tt}E^{2}+g^{\phi\phi}L_{z}^{2}=E^{2}(g^{tt}+g^{\phi\phi}\eta^{2}).
\end{eqnarray}
$E$ and $L_{z}$ are two constants of motion for the null geodesics motion, i.e., energy and the $z$-component of the angular momentum, so the impact parameter $\eta=L_{z}/E$ is also a constant for the photon motion. The boundary of shadow is determined by the photon sphere composed by the unstable photon circular orbits. The light rays that enter the photon sphere will be captured by black hole; the light rays that do not enter the photon sphere will fly away to infinity; the light rays that spiral asymptotically towards the photon sphere will compose the boundary of black hole shadow. Now, let us study the spherical photon orbits in the equatorial plane also known as light rings. The light rings must satisfy
\begin{eqnarray}
\label{lr}
\theta=\pi/2,\;\;\;\;\;\;\;\; V_{eff}=0,\;\;\;\;\;\;\;\;\frac{\partial V_{eff}}{\partial r}=0.
\end{eqnarray}
Moreover, the light ring with $\partial^{2}V_{eff}/\partial r^{2}<0$ is unstable, the light ring with $\partial^{2}V_{eff}/\partial r^{2}>0$ is stable. Solving the equations (\ref{lr}), we found the radius of light rings $r_{LR}$ only depends on the quadrupole strength $\mathcal{Q}$, shown in Fig.\ref{rlq}. One can find there is a critical value of the quadrupole strength $\mathcal{Q}_{c}\approx-0.0209445$ for light ring in Fig.\ref{rlq}. The light ring doesn't exist in the space-time of Schwarzschild black hole with halo for $\mathcal{Q}<\mathcal{Q}_{c}$; both the unstable (red dashed line) and stable (black line) light rings exist for $\mathcal{Q}_{c}<\mathcal{Q}<0$; only one unstable (red dashed line) light ring exists for $\mathcal{Q}>0$. Our previous work\cite{scc} manifested that the existence of stable light ring will make the photons in stable orbits are always moving around the black hole, cannot enter the black hole or escape to infinity. The inexistence of light ring will make a panoramic (equatorial) shadow appear\cite{scc,schm}. In this paper, we only research the case that the light ring exist and the halo structures could be considered as a perturbation of a black hole. After research we prefer to set the quadrupole strength $\mathcal{Q}$ on the order of $10^{-4}$, and to set the quadrupole strength $\mathcal{O}$ on the order of $10^{-6}$. In this case, the halo structures are more in line with the actual astronomical situation. Moreover, the radius of stable light ring are bigger than the radius $r_{obs}$ of the observer we set (for $\mathcal{Q}=-1\times10^{-4}$, $r_{LR}=70.7M>r_{obs}=50M$), in which the observer will directly observe the black hole shadow.

\section{The shadow casted by Schwarzschild black hole with halo}
Now, we study the shadow of Schwarzschild black hole with halo through the backward ray-tracing technique \cite{sw,swo,astro,chaotic,my,sha18,sMN}. In this method, we assume that the static observer is locally at ($r_{obs}, \theta_{obs}$) in zero-angular-moment-observers (ZAMOs) reference frame \cite{sha2}, and evolved light rays from the observer backward in time. The shadow of black hole is composed by the light rays falling down into the event horizon of black hole. The coordinates of a photon's image in observer's sky can be expressed as \cite{sw,swo,astro,chaotic,my,sha18,sMN}
\begin{eqnarray}
\label{xd1}
x&=&-r_{obs}\frac{\sqrt{1-\frac{2M}{r}}L_{z}}{r\sin\theta e^{(\frac{1}{2}\lambda_{Q}+\frac{1}{2}\lambda_{O}+\frac{1}{2}\lambda_{QO}-\nu_{Q}-\nu_{O})}\dot{r}}|_{(r_{obs},\theta_{obs})}, \nonumber\\
y&=&r_{obs}\frac{\sqrt{r(r-2M)}\dot{\theta}}{\dot{r}}|_{(r_{obs},\theta_{obs})}.
\end{eqnarray}

\begin{figure}
\includegraphics[width=16.5cm ]{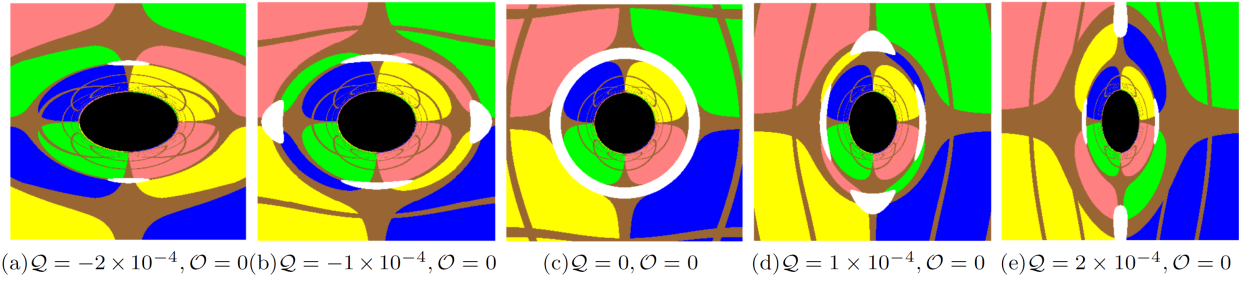}
\caption{The shadows of Schwarzschild black hole with the halo for the quadrupole strength $\mathcal{Q}=-2\times10^{-4}, -1\times10^{-4}, 0, 1\times10^{-4}, 2\times10^{-4}$, and the octopolar strength $\mathcal{O}=0$. Here we set $M=1$ and the static observer at $r_{obs}=50$ with the inclination angle $\theta_{obs}=\pi/2$.}
\label{q}
\end{figure}

\begin{figure}
\includegraphics[width=16.5cm ]{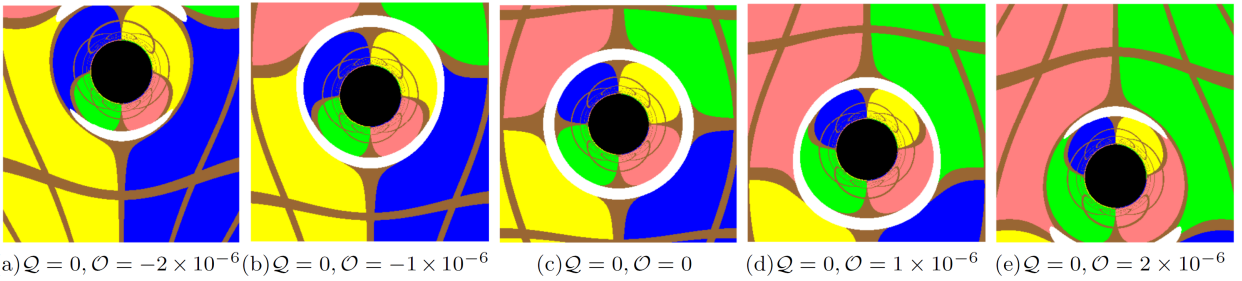}
\caption{The shadows of Schwarzschild black hole with the halo for the quadrupole strength $\mathcal{Q}=0$, and the octopolar strength $\mathcal{O}=-2\times10^{-6}, -1\times10^{-6}, 0, 1\times10^{-6}, 2\times10^{-6}$. Here we set $M=1$ and the static observer at $r_{obs}=50$ with the inclination angle $\theta_{obs}=\pi/2$.}
\label{o}
\end{figure}

\begin{figure}
\includegraphics[width=16.5cm ]{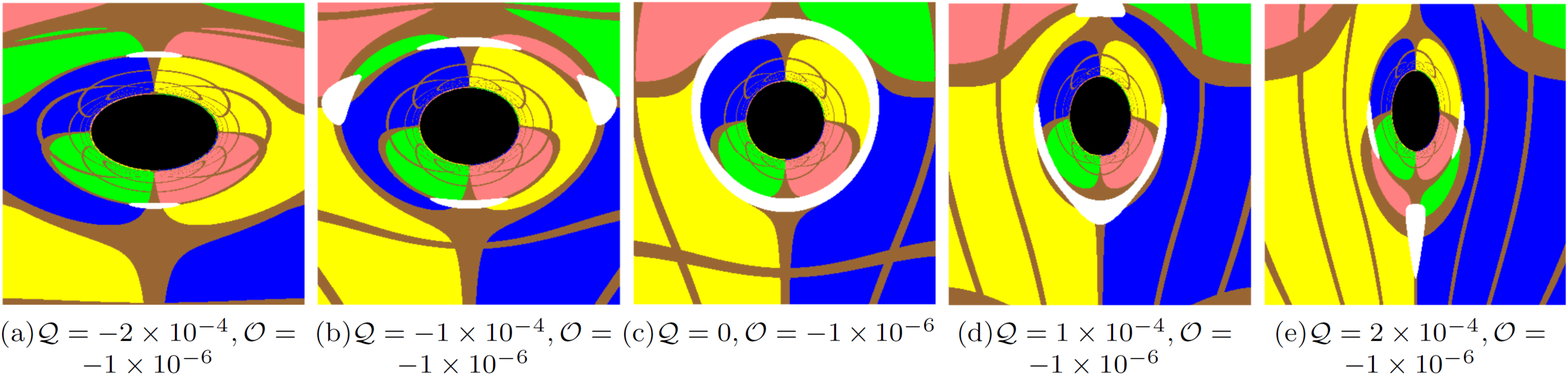}
\caption{The shadows of Schwarzschild black hole with the halo for the quadrupole strength $\mathcal{Q}=-2\times10^{-4}, -1\times10^{-4}, 0, 1\times10^{-4}, 2\times10^{-4}$, and the octopolar strength $\mathcal{O}=-1\times10^{-6}$. Here we set $M=1$ and the static observer at $r_{obs}=50$ with the inclination angle $\theta_{obs}=\pi/2$.}
\label{qo}
\end{figure}

\begin{figure}
\includegraphics[width=16.5cm ]{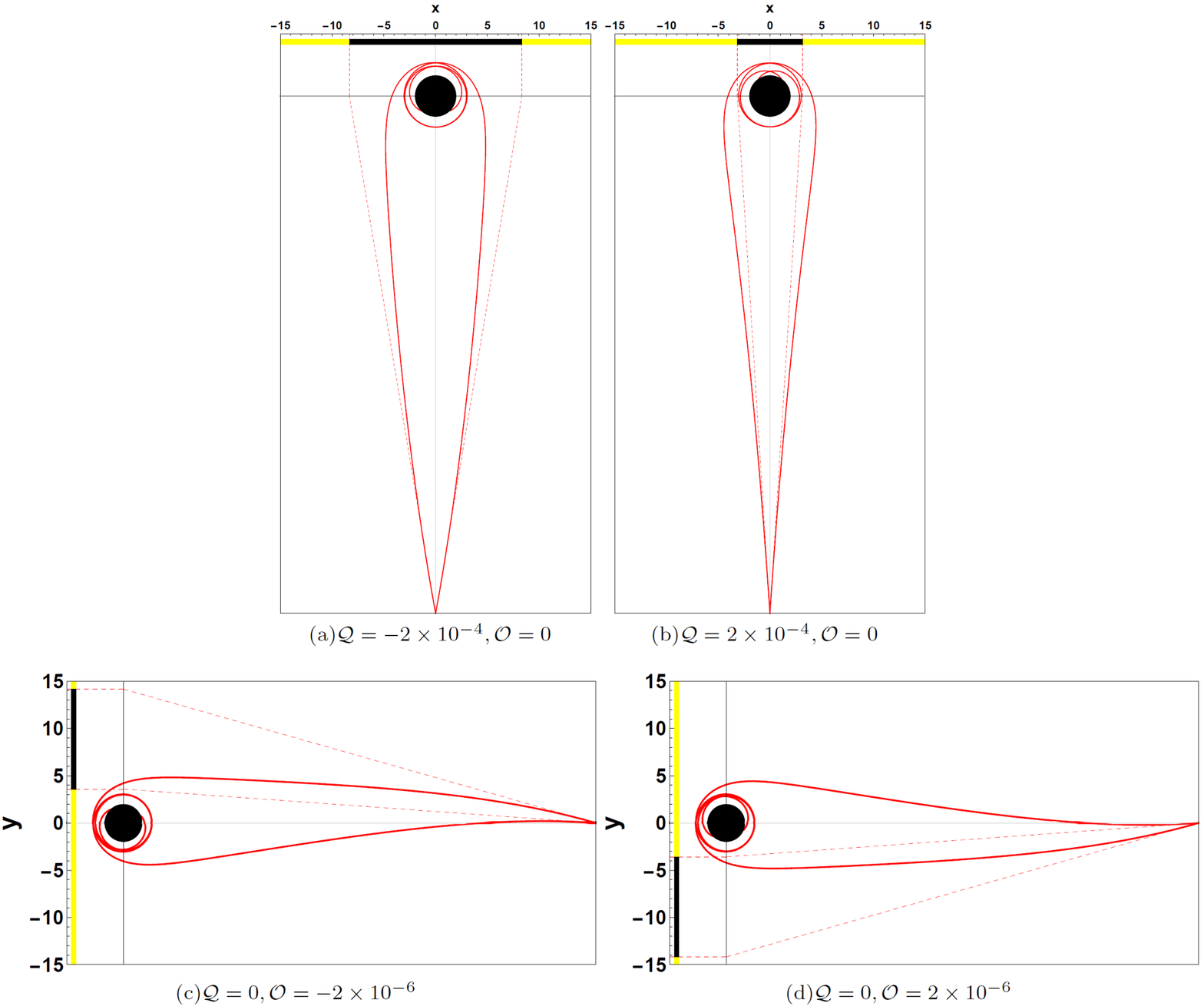}
\caption{The light rays (red lines) spiraling asymptotically towards the photon sphere and their tangents (red dash lines) at the observer that determine the boundary of black hole shadow. Black hole shadow stretches and squeezes on the celestial sphere for $\mathcal{Q}<0$ and $\mathcal{Q}>0$ respectively, and shifts upward for $\mathcal{O}<0$ and shifts downward for $\mathcal{O}>0$.}
\label{gjqo}
\end{figure}

In Fig.\ref{q} we show the influence of the quadrupole term on Schwarzschild black hole shadow with different quadrupole strength $\mathcal{Q}$. Here we set $M=1$ and the static observer at $r_{obs}=50$ with the inclination angle $\theta_{obs}=\pi/2$. In this paper, we set a light-emitting celestial sphere as background light source marked by four different colored quadrants and brown grids as longitude and latitude, which is same as the celestial sphere in Ref.\cite{my,sha18,sw}. The shadow of Schwarzschild black hole only with the quadrupole structure is symmetric about the equatorial plane. One can find the shadow of Schwarzschild black hole with halo becomes more oblate by only stretching along the horizontal direction as $\mathcal{Q}$ decreases for $\mathcal{Q}<0$. But for $\mathcal{Q}>0$, the black hole shadow becomes more prolate by only squeezing along the horizontal direction as $\mathcal{Q}$ increases. With the increase of $|\mathcal{Q}|$, the Einstein ring, the white ring around black hole shadow, is torn into an Einstein cross. In Fig.\ref{o} we show the influence of the octopolar term on Schwarzschild black hole shadow with different octopolar strength $\mathcal{O}$. One can find the black hole shadow shifts upward as $\mathcal{O}$ decreases for $\mathcal{O}<0$, but the shadow shifts downward as $\mathcal{O}$ increases for $\mathcal{O}>0$. The octopolar term also breaks the reflection symmetry of Schwarzschild black hole shadow with respect equatorial plane, but the two black hole shadows with the opposite $\mathcal{O}$ are symmetrical to each other about the equatorial plane. In Fig.\ref{qo} we show joint efforts of the quadrupole and octopole terms on the shadow of Schwarzschild black hole. That is the black hole shadow stretches and squeezes along the horizontal direction for $\mathcal{Q}<0$ and $\mathcal{Q}>0$ respectively, and black hole shadow shifts upward for $\mathcal{O}<0$ and shifts downward for $\mathcal{O}>0$. In Fig.\ref{gjqo} we exhibit the light rays that form the shadow boundary to explain the emergence of the extraordinary patterns of black hole shadow with the quadrupole and octopole terms. Fig.\ref{gjqo}(a) and (b) show the light rays with $\mathcal{Q}=-2\times10^{-4}$ and $\mathcal{Q}=2\times10^{-4}$ respectively in the plane $y=0$ (the equatorial plane), where $y$ is the celestial coordinate(\ref{xd1}) in observer's sky. One can find the two light rays (red lines) propagate from the observer backward in time, and spiral asymptotically towards the photon sphere. But for the observer, the light rays propagate along the red dash lines, and determine the left and right margin of black hole shadow respectively. The black region in the $x$ axis represent the black hole shadow on the celestial sphere, where $x$ is the celestial coordinate(\ref{xd1}). One can find the angle between the two red dash lines, namely the angular radius of black hole shadow, is much larger for $\mathcal{Q}=-2\times10^{-4}$, and is much smaller for $\mathcal{Q}=2\times10^{-4}$. Which results in the stretching and squeezing of black hole shadow in the horizontal direction for $\mathcal{Q}<0$ and $\mathcal{Q}>0$ respectively. Fig.\ref{gjqo}(c) and (d) show the light rays with $\mathcal{O}=-2\times10^{-6}$ and $\mathcal{O}=2\times10^{-6}$ respectively in the plane $x=0$. We also exhibit the light rays (red lines) spiraling asymptotically towards the photon sphere and their tangents (red dash lines) at the observer that determine the upper and lower margin of shadow. One can find the tangents slope upward for $\mathcal{O}=-2\times10^{-6}$, and slope downward for $\mathcal{O}=2\times10^{-6}$. Which causes the shadow to shift upward and downward in the $y$ axis of celestial sphere for $\mathcal{O}<0$ and $\mathcal{O}>0$ respectively.


\begin{figure}[htb]
\includegraphics[width=8cm ]{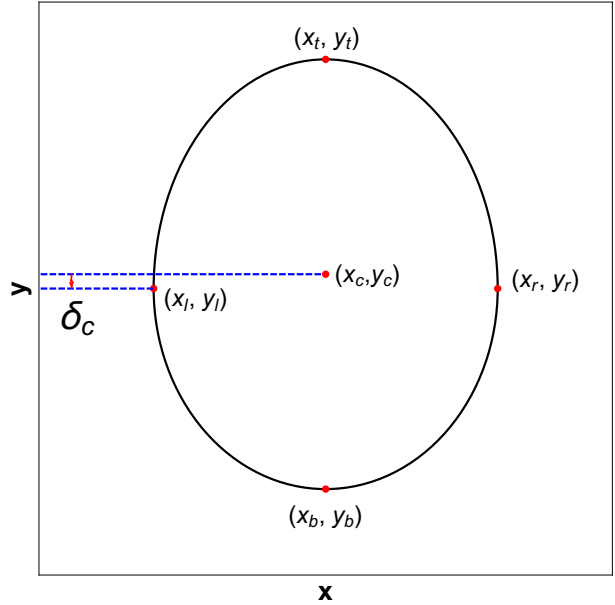}
\caption{The leftmost point ($x_{l}$, $y_{l}$), the rightmost point ($x_{r}$, $y_{r}$), the topmost point ($x_{t}$, $y_{t}$), the bottommost point ($x_{b}$, $y_{b}$) and the center $(x_{c}, y_{c})=(\frac{x_{l}+x_{r}}{2},\frac{y_{t}+y_{b}}{2})$ of black hole shadow. The distortion parameter $\delta_{c}$ is defined as $(y_{c}-y_{l})/R_{s}$, where $R_{s}$ is the radius of Schwarzschild black hole shadow ($\mathcal{Q}=\mathcal{O}=0$).}
\label{csd}
\end{figure}

\begin{figure}
\includegraphics[width=16cm ]{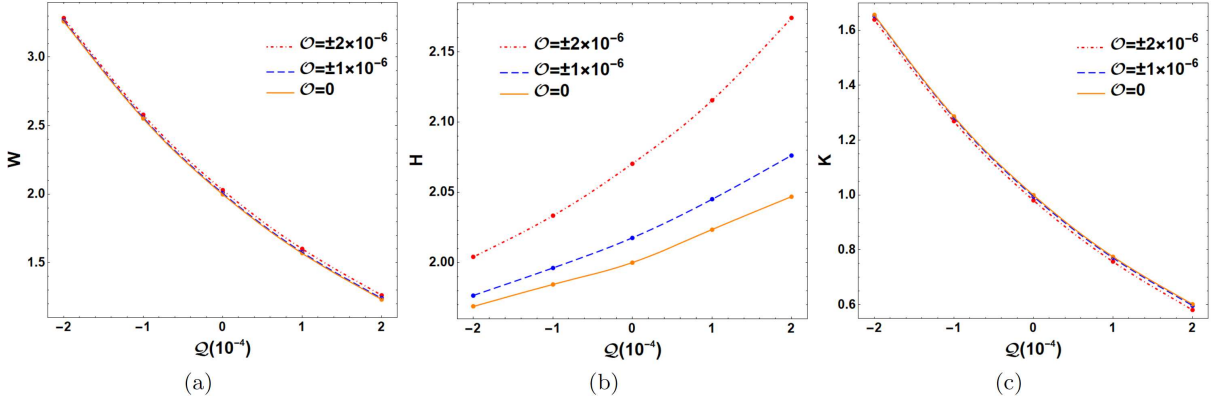}
\caption{The varieties of the width $W$, the height $H$ and the oblateness $K$ of black hole shadow with the quadrupole strength $\mathcal{Q}$ for different $\mathcal{O}$.}
\label{whk}
\end{figure}

\begin{figure}
\includegraphics[width=15cm ]{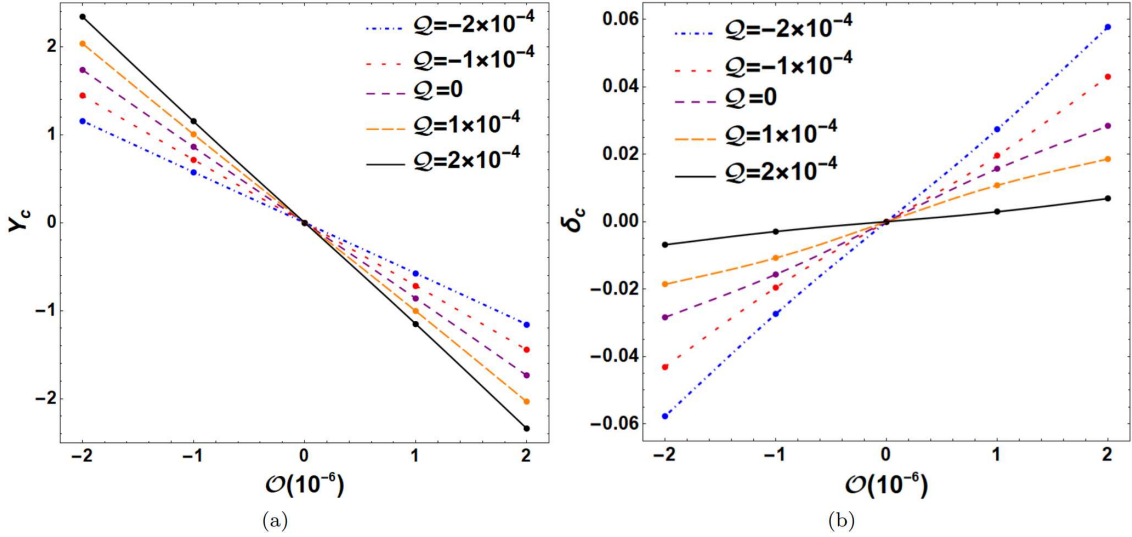}
\caption{The varieties of the deviation $Y_{c}$ and the distortion parameter $\delta_{c}$ of black hole shadow with the octopolar strength $\mathcal{O}$ for different $\mathcal{Q}$.}
\label{yc}
\end{figure}

Black hole in the universe could be perturbed by a halo with quadrupole and octopolar terms, so we estimated several observables of black hole shadow, hope to determine the quadrupole strength $\mathcal{Q}$ and octopolar strength $\mathcal{O}$ in astronomical observations. To characterize the shadow of Schwarzschild black hole with halo, we should first introduce four important points for shadow: the leftmost point ($x_{l}$, $y_{l}$), the rightmost point ($x_{r}$, $y_{r}$), the topmost point ($x_{t}$, $y_{t}$) and the bottommost point ($x_{b}$, $y_{b}$), shown in Fig.\ref{csd}. So we can define black hole shadow's observable values: the width $W=(x_{r}-x_{l})/R_{s}$, the height $H=(y_{t}-y_{b})/R_{s}$ and the oblateness $K=W/H$, where $R_{s}$ is the radius of Schwarzschild black hole shadow($\mathcal{Q}=\mathcal{O}=0$). Fig.\ref{whk} shows the varieties of the width $W$, the height $H$ and the oblateness $K$ of black hole shadow with the quadrupole strength $\mathcal{Q}$ for different octopolar strength $\mathcal{O}$. In Fig.\ref{whk}(a), one can find the width $W$ of black hole shadow decreases as $\mathcal{Q}$ increases, and almost all the width $W$ are larger than $2R_{s}$ for $\mathcal{Q}<0$, are less than $2R_{s}$ for $\mathcal{Q}>0$. Moreover, the octopolar strength $\mathcal{O}$ barely has effect on the width $W$ of black hole shadow. It indicates that only the quadrupole term stretches and squeezes black hole shadow along the horizontal direction for $\mathcal{Q}<0$ and $\mathcal{Q}>0$ respectively. In Fig.\ref{whk}(b), one can find the height $H$ of black hole shadow increases as $\mathcal{Q}$ increases, but the change in height is much small comparing with the change in width. Thus the change of oblateness $K$ of black hole shadow with $\mathcal{Q}$ is almost the same to the change of width $W$. The black hole shadow is oblate ($K>1$) for $\mathcal{Q}<0$ and is prolate ($K<1$) for $\mathcal{Q}>0$. In addition, the height $H$ of black hole shadow increases as $|\mathcal{O}|$ increases for the fixed $\mathcal{Q}$. But the main effect of the octopolar strength $\mathcal{O}$ is to make black hole shadow move along the vertical direction. So we define the center of black hole shadow as $(x_{c}, y_{c})=(\frac{x_{l}+x_{r}}{2},\frac{y_{t}+y_{b}}{2})$, and make use of $Y_{c}$($y_{c}/R_{s}$) to describe the deviation of shadow away from the equatorial plane. In Fig.\ref{yc}(a), we show the varieties of $Y_{c}$ with the octopolar strength $\mathcal{O}$ for different $\mathcal{Q}$. One can find the deviation $|Y_{c}|$ increases as $|\mathcal{O}|$ increases, and $Y_{c}>0$ for $\mathcal{O}<0$, $Y_{c}<0$ for $\mathcal{O}>0$. What's more, the deviation $|Y_{c}|$ is bigger for bigger $\mathcal{Q}$. It indicates the prolate shadow will amplify the deviation of shadow and the oblate shadow will minish the deviation. Unfortunately, in the actual observations of black hole shadow we can't determine the coordinates of shadow center $(x_{c}, y_{c})$ to estimate the value of $\mathcal{O}$. But the octopolar term, meanwhile, can cause a slight distortion along the vertical direction in the black hole shadow that is $y_{c}$ isn't equal to $y_{l}$ or $y_{r}$, shown in Fig.\ref{csd}. So we can define a distortion parameter $\delta_{c}=(y_{c}-y_{l})/R_{s}$ to describe the distortion caused by the octopolar term. Fig.\ref{yc}(b) shows the varieties of the distortion parameter $\delta_{c}$ with the octopolar strength $\mathcal{O}$ for different $\mathcal{Q}$. One can find the distortion parameter $|\delta_{c}|$ increases as $|\mathcal{O}|$ increases, and $\delta_{c}<0$ for $\mathcal{O}<0$, $\delta_{c}>0$ for $\mathcal{O}>0$. In addition, the distortion parameter $|\delta_{c}|$ is smaller for bigger $\mathcal{Q}$. It indicates the prolate shadow will minish the distortion and the oblate shadow will amplify the distortion. We hope Event Horizon Telescope and BlackHoleCam could observe the shadow of a black hole perturbed by a halo with quadrupole and octopolar terms in the future astronomical observations.


\begin{figure}
\includegraphics[width=15cm ]{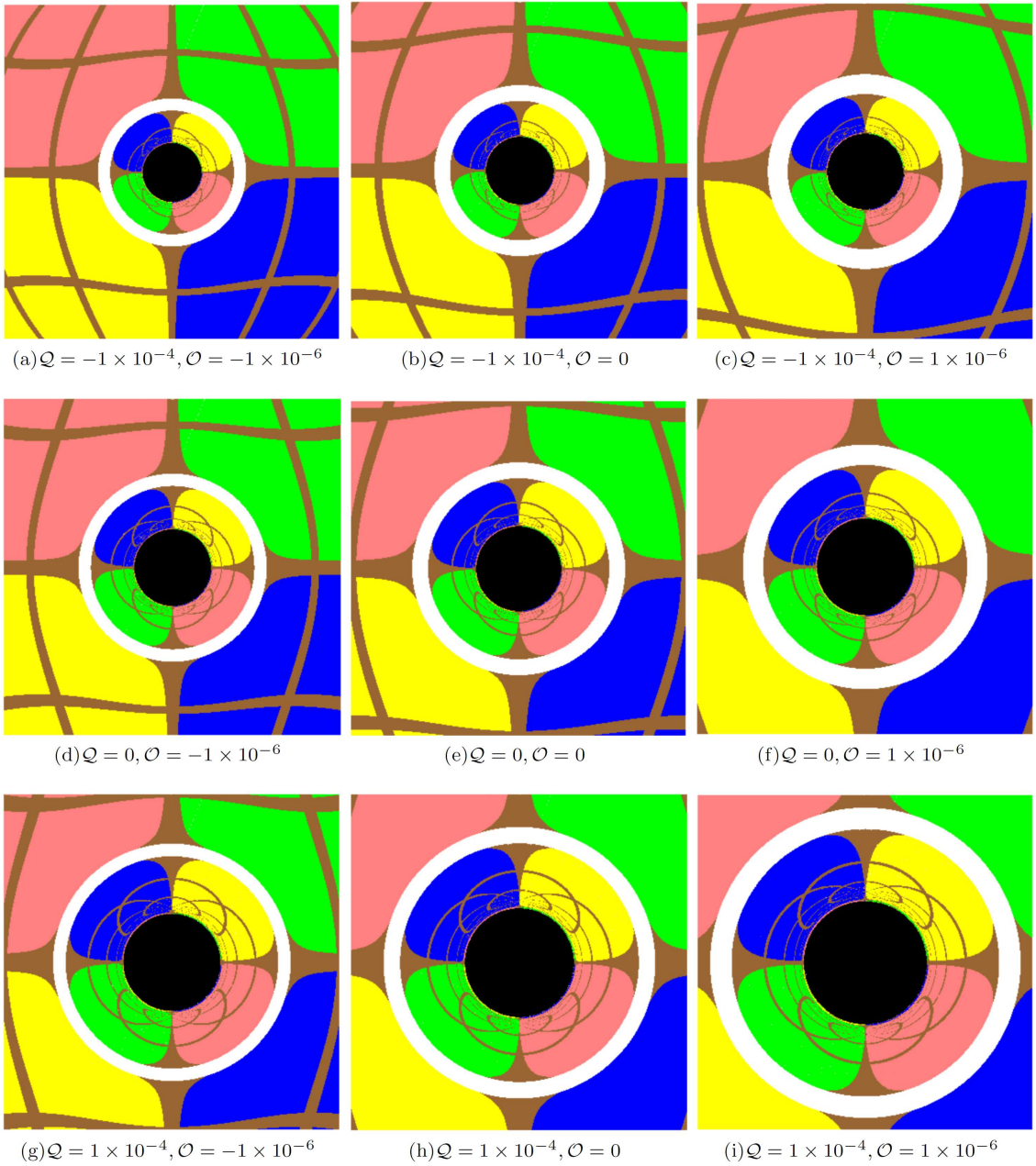}
\caption{The shadows of Schwarzschild black hole with halo for the observer with the inclination angle $\theta_{obs}=0$. Top row: the quadrupole strength $\mathcal{Q}=-1\times10^{-4}$, the octopolar strength $\mathcal{O}=-1\times10^{-6}, 0, 1\times10^{-6}$. Middle row: $\mathcal{Q}=0$, $\mathcal{O}=-1\times10^{-6}, 0, 1\times10^{-6}$. Bottom row: $\mathcal{Q}=1\times10^{-4}$, $\mathcal{O}=-1\times10^{-6}, 0, 1\times10^{-6}$.}
\label{0d}
\end{figure}

\begin{figure}
\includegraphics[width=15cm ]{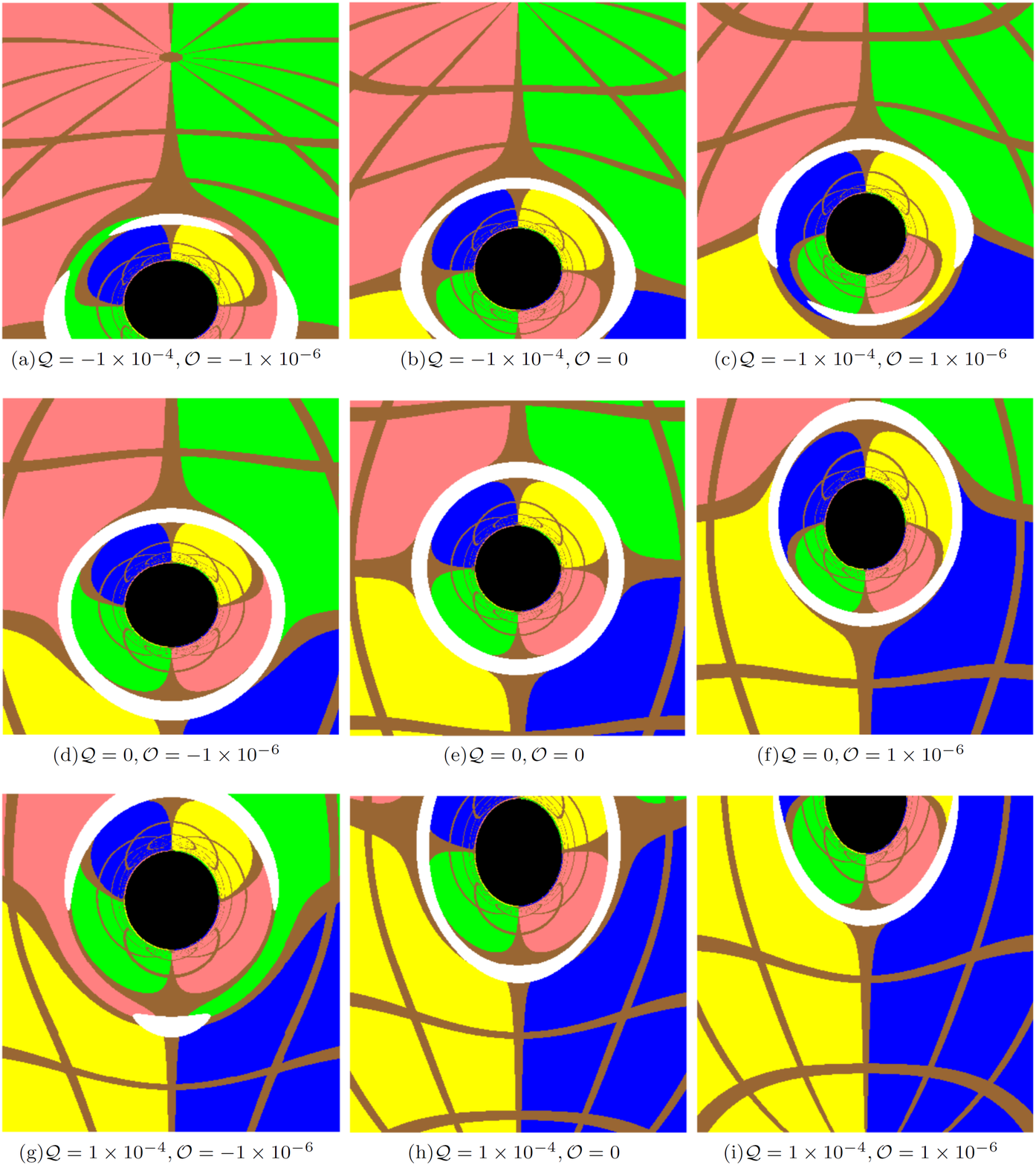}
\caption{The shadows of Schwarzschild black hole with halo for the observer with the inclination angle $\theta_{obs}=\pi/4$. Top row: the quadrupole strength $\mathcal{Q}=-1\times10^{-4}$, the octopolar strength $\mathcal{O}=-1\times10^{-6}, 0, 1\times10^{-6}$. Middle row: $\mathcal{Q}=0$, $\mathcal{O}=-1\times10^{-6}, 0, 1\times10^{-6}$. Bottom row: $\mathcal{Q}=1\times10^{-4}$, $\mathcal{O}=-1\times10^{-6}, 0, 1\times10^{-6}$.}
\label{45d}
\end{figure}

In Figs.\ref{0d} and \ref{45d}, we present the shadows of Schwarzschild black hole with halo for the observer inclination angle $\theta_{obs}=0$ and $\pi/4$ respectively. From Fig.\ref{0d} one can find black hole shadows are always circle with different $\mathcal{Q}$ and $\mathcal{O}$ for $\theta_{obs}=0$. But the interesting thing is black hole shadow becomes bigger with the increase of $\mathcal{Q}$ or $\mathcal{O}$. Comparing to Schwarzschild black hole shadow (Figs.\ref{0d}(e) with $\mathcal{Q}=\mathcal{O}=0$), one can find the negative $\mathcal{Q}$ or negative $\mathcal{O}$ will make black hole shadow shrink, and the positive $\mathcal{Q}$ or positive $\mathcal{O}$ will make black hole shadow expand. For the observer inclination angle $\theta_{obs}=\pi/4$(Fig.\ref{45d}), black hole shadow not only becomes more prolate but also shifts upward with the increase of $\mathcal{Q}$ or $\mathcal{O}$.

\section{summary}

We have studied the surface geometry and the shadows of Schwarzschild black hole with halo. The exterior halo is a multipolar structure containing quadrupolar and octopolar terms. We found the quadrupole term makes the Schwarzschild black hole prolate for the quadrupole strength $\mathcal{Q}<0$ and oblate for $\mathcal{Q}>0$, and the octopole term makes shadow stretch upward for the octopolar strength $\mathcal{O}<0$ and stretch downward for $\mathcal{O}>0$. The radius of light rings $r_{LR}$ only depends on $\mathcal{Q}$ in the space-time of Schwarzschild black hole with halo. The light ring doesn't exist when $\mathcal{Q}$ is larger than a critical $\mathcal{Q}_{c}$; both unstable and stable light rings exist for $\mathcal{Q}_{c}<\mathcal{Q}<0$; only one unstable light ring exists for $\mathcal{Q}>0$. The shadow of Schwarzschild black hole with halo stretches and squeezes along the horizontal direction for $\mathcal{Q}<0$ and $\mathcal{Q}>0$ respectively. Meanwhile, black hole shadow shifts upward for $\mathcal{O}<0$ and shifts downward for $\mathcal{O}>0$. We exhibited the light rays spiraling asymptotically towards the photon sphere and their tangents at the observer that determine the margin of shadow. The angle between the two tangents, namely the angular radius of black hole shadow, is larger for $\mathcal{Q}<0$, and is smaller for $\mathcal{Q}>0$, which results in the stretching and squeezing of black hole shadow in the horizontal direction. The tangents slope upward for $\mathcal{O}<0$, and slope downwards for $\mathcal{O}>$, which causes the shadow to shift upward and downward in the vertical direction. From the observable width $W$, height $H$, oblateness $K$ and distortion parameter $\delta_{c}$ of black hole shadow, one can determine the value of $\mathcal{Q}$ and $\mathcal{O}$ of Schwarzschild black hole with halo. Black hole shadow is always a circle and becomes bigger as $\mathcal{Q}$ or $\mathcal{O}$ increases for the observer inclination angle $\theta_{obs}=0$. Black hole shadow not only becomes more prolate but also shifts upward with the increase of $\mathcal{Q}$ or $\mathcal{O}$ for $\theta_{obs}=\pi/4$. Our results show that the quadrupolar and octopolar terms yield a series of interesting patterns for the shadow of a Schwarzschild black hole with halo.

\section{\bf Acknowledgments}

This work was supported by the National Natural Science Foundation of China under Grant No. 12105151, the Shandong Provincial Natural Science Foundation of China under Grant No. ZR2020QA080, and was partially supported by the National Natural Science Foundation of China under Grant No. 11875026, 11875025 and 12035005.

\end{document}